# Current Progress of Digital Twin Construction Using Medical Imaging

Feng Zhao[a,b], Yizhou Wu[a,c], Mingzhe Hu[a,d], Chih-Wei Chang[a], Ruirui Liu[e], Richard Qiu[a] and Xiaofeng Yang[a,d]

[a]*Department of Radiation Oncology and Winship Cancer Institute, Emory University, Atlanta, GA 30322, USA*
[b]*School of Computational Science and Engineering, Georgia Institute of Technology, GA, Atlanta, USA*
[c]*School of Electrical and Computer Engineering, Georgia Institute of Technology, GA, Atlanta, USA*
[d]*Department of Computer Science and Informatics, Emory University, GA, Atlanta, USA*
[e]*Department of Radiation Oncology, University of Nebraska Medical Center, NE, Omaha, USA*
*Email: xiaofeng.yang@emory.edu*

---

**Abstract**

Medical imaging has played a pivotal role in advancing and refining digital twin technology, allowing for the development of highly personalized virtual models that represent human anatomy and physiological functions. A key component in constructing these digital twins is the integration of high-resolution imaging data—such as MRI, CT, PET, and ultrasound—with sophisticated computational models. Advances in medical imaging significantly enhance real-time simulation, predictive modeling, and early disease diagnosis, individualized treatment planning, ultimately boosting precision and personalized care. Although challenges persist, such as the complexity of anatomical modeling, integrating various imaging modalities, and high computational demands, recent progress in imaging and machine learning has greatly improved the precision and clinical applicability of digital twins. This review investigates the role of medical imaging in developing digital twins across organ systems. Key findings demonstrate that improvements in medical imaging have enhanced the diagnostic and therapeutic potential of digital twins beyond traditional methods, particularly in imaging accuracy, treatment effectiveness, and patient outcomes. The review also examines the technical barriers that currently limit further development of digital twin technology, despite advances in medical imaging, and outlines future research avenues aimed at overcoming these challenges to unlock the full potential of this technology in precision medicine.

**Table 1.** Technical abbreviations.

| Full Term | Abbreviation | Full Term | Abbreviation |
|---|---|---|---|
| Abdominal Aortic Aneurysm | AAA | High-performance Computing | HPC |
| Alzheimer's Disease | AD | Image Reconstruction Network | IR-Net |
| Aortic Stenosis | AS | Implantable Cardioverter Defibrillator | ICD |
| Artificial Intelligence | AI | Internet of Things | IoT |
| Atrial Fibrillation | AF | Intrahepatic Portosystemic Shunts | IPSS |
| Augmented Reality | AR | Inverse Kinematics | IK |
| Blood Oxygen Level Dependent | BOLD | Level Set Method | LSM |
| Cardiac Resynchronization Therapy | CRT | Linear Regression | LR |
| Common Objects in Context | COCO | Long Short-Term Memory | LSTM |
| Computational Fluid Dynamics | CFD | Magnetic Resonance | MR |
| Computed Tomography | CT | Magnetic Resonance Imaging | MRI |
| Computed Tomography Angiography | CTA | Magnetic Resonance Spectroscopy | MRS |
| Computer-Aided Manufacturing | CAM | Metabolic Dysfunction-Associated Fatty Liver Disease | MAFLD |
| Cone Beam Computed Tomography | CBCT | Mixture of Products of the Multinomials | MPoM |
| Convolutional Neural Network | CNN | Multi-Core Fiber | MCF |
| Central Nervous System | CNS | Multiple Sclerosis | MS |
| Coronary Artery Disease | CAD | Neoadjuvant Systemic Therapy | NAST |
| Decision Tree Regression | DTR | Neural Ordinary Differential Equations | NeuralODEs |
| Deep Convolutional Generative Adversarial Network | DCGAN | Nonalcoholic Fatty Liver Disease | NAFLD |
| Deep Convolutional Neural Network | DCNN | Optical Coherence Tomography | OCT |
| Deep Neural Network | DNN | Peak Signal-to-Noise Ratio | PSNR |
| Dice Similarity Coefficient | DSC | Portal Hypertension | PHT |
| Diffusion Tensor Imaging | DTI | Principal Component Analysis | PCA |
| Diffusion-Weighted MRI | DW-MRI | Proton Density Fat Fraction | PDFF |
| Digital Imaging and Communications in Medicine | DICOM | Preferred Reporting Items for Systematic Reviews and Meta-Analyses | PRISMA |
| Digital Twin Brain | DTB | Random Forest Regression | RFR |
| Digital Twin Brain Visualization | DTBVis | Recurrent Neural Network | RNN |
| Discrete Multiphysics | DMP | Reduced-Order Model | ROM |
| Doppler Ultrasound | DUS | Relative Technical Error of Measurement | rTEM |
| Double-Outlet Right Ventricle | DORV | Robotic-assisted Radical Prostatectomy | RARP |
| Dynamic Colon Model | DCM | Robust Auxiliary Classifier GAN | rAC-GAN |
| Dynamic Contrast-Enhanced Magnetic Resonance Imaging | DCE-MRI | Semi-Supervised Support Vector Machines | S3VM |
| Electrical Impedance Tomography | EIT | Smoothed Particle Hydrodynamics | SPH |
| Electrocardiogram | ECG | Standard Tessellation Language | STL |
| Electromyography | EMG | Statistical Shape Model | SSM |
| Electrophysiology | EP | Structural Similarity Index Measure | SSIM |
| Enhanced Deep Super-Resolution | EDSR | Sudden Cardiac Death | SCD |
| Extended Reality | XR | Support Vector Machine | SVM |
| Finite Element Modeling | FEM | t-distributed Stochastic Neighbor Embedding | t-SNE |
| Four-Dimensional Computed Tomography | 4D-CT | Three-Dimensional | 3D |
| Fully Convolutional Network | FCN | Three-Dimensional Virtual Modeling | 3DVM |
| Finite Element Analysis | FEA | Triple-Negative Breast Cancer | TNBC |
| Functional Connectivity Matrices | FC matrices | Two-Dimensional | 2D |
| functional MRI | fMRI | Ventricular Tachycardia | VT |
| Gaussian Process Regression | GPR | Verification, Validation, and Uncertainty | VVUQ |
| | | Quantification | |
| Geographic Atrophy | GA | You Only Look Once | YOLO |
| Generative Artificial Intelligence | GAI | Zero-Dimensional | 0D |
| Generative Adversarial Networks | GANs | | |
| Gradient Boosting Algorithm | GBA | | |
| Graph Neural Network | GNN | | |
| Graphics Processing Unit | GPU | | |
| Heart Failure | HF | | |
| Hounsfield Unit | HU | | |
| Hybrid Pyramid U-Net | HPU-Net | | |
| Hyperbolic Graph Neural Networks | HGNN | | |
| Hypergraph Convolutional Neural Network | HGCNN | | |

# 1. Introduction

Originally emerging in the manufacturing and aerospace sectors (Glaessgen and Stargel, 2012), digital twin technology has now entered healthcare, transforming the field of medical imaging. A digital twin is a dynamic, virtual representation of a physical entity, created using real-time data and sophisticated computational models. With advancements in medical imaging, digital twins allow for the creation of highly detailed and precise representations of a patient's anatomy and physiological processes. One significant application of digital twins in healthcare is their ability to support personalized treatment for each patient, with the added advantage of dynamically updating the model based on longitudinal measurements and clinical data, thereby capturing the patient's disease trajectory over time. Digital twins enable healthcare professionals to forecast disease progression, predict responses to interventions such as drugs, and anticipate adverse events associated with treatments. As a result, there is an increasing need for advanced tools and approaches to meet these demands effectively. Digital twin technology is emerging as a powerful tool in healthcare, opening new possibilities for enhancing patient outcomes, optimizing treatment plans, and advancing preventive care strategies.

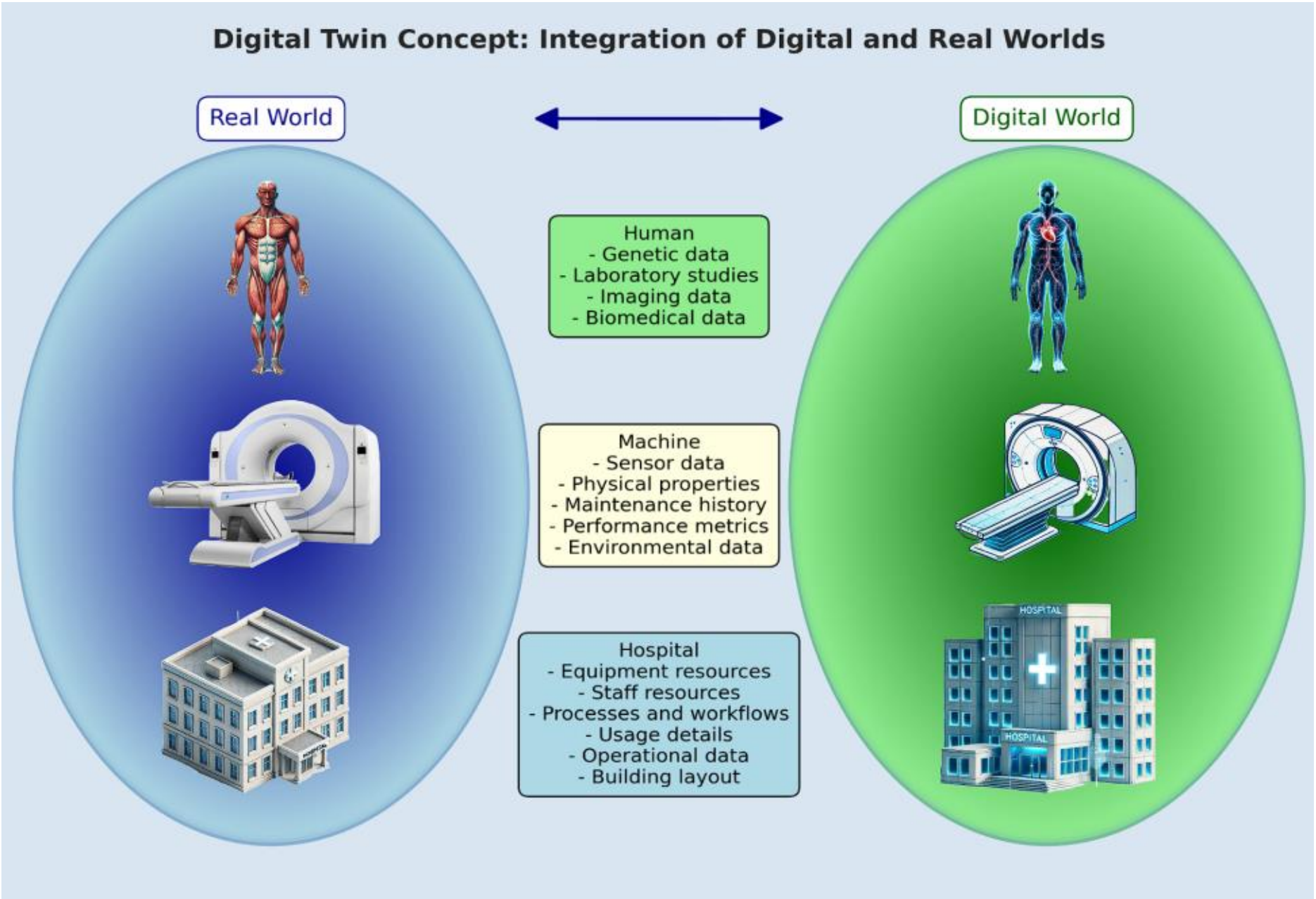

**Figure 1**. Comprehensive integration of digital twin technology in healthcare.

Digital twin technology is revolutionizing personalized healthcare across various medical domains. For instance, a cardiovascular digital-twin system (Trayanova and Prakosa, 2024) had been integrated using MRI and CT with hemodynamic data to simulate heart and blood flow dynamics, aiding in the prediction of arrhythmias or blockages and facilitating precise intervention planning. Figure 1 illustrates the comprehensive integration of digital twin technology in healthcare. On the left side, representing the physical world, are key entities: the human body, medical imaging devices, and healthcare facilities. Data is continuously collected from these sources, including genetic information, lab results, imaging data, and other biomedical data from patients; sensor readings, physical properties, and performance metrics from medical devices; and resource usage, staff allocation, processes, and workflows from hospitals. This real-world data is then transferred to the digital realm, depicted on the right side of Figure 1. Digital twins of the human body, medical devices, and healthcare facilities are created and regularly updated. The digital twin of a human aggregates all the collected data, forming a dynamic virtual representation that enables detailed analysis and simulation of patient-specific conditions. Similarly, digital twins of machines and hospital systems incorporate operational and environmental data, providing a complete view of their performance and usage. Together, these digital twins offer an integrated perspective on healthcare operations, enhancing patient care, resource management, and overall efficiency.

Medical imaging is fundamental to constructing digital twins, serving as the critical link between the real patient and their virtual representation. While these models are continuously enhanced with data from multiple sources, imaging technologies such as MRI, CT, and PET provide the essential, non-invasive, high-resolution, and dynamic data needed for precise digital replication. By capturing detailed anatomical structures, identifying abnormalities, and documenting physiological processes, these imaging modalities enable ongoing monitoring and simulation of disease progression and treatment responses. Beyond simple visualization, medical imaging provides deep anatomical and functional insights that reveal each patient's unique characteristics. Utilizing these advanced imaging technologies addresses the complexities of replicating the intricate human body within a virtual environment, driving forward the possibilities in personalized medicine.

The evolution of digital twin technology has reached a pivotal stage in precision medicine, driven by improvements in computational power, high-performance computing, and cloud-based platforms that facilitate complex simulations. This progress has been further accelerated by the development of deep learning models, particularly foundational models, which overcome significant limitations of traditional analytical or physical models. This is especially valuable in complex systems like the human circulatory system, where many physiological parameters are difficult or impossible to measure directly, complicating accurate modeling. Foundational models, pre-trained on extensive datasets, have made the construction of digital twins more feasible by enabling robust modeling even with incomplete or diverse data. Moreover, the integration of multimodal data—including medical imaging, genomic information, wearable sensor data, and electronic health records—enables a comprehensive and dynamic representation of individual patients. This integration supports real-time monitoring, predictive modeling, and personalized treatment planning. Machine learning techniques, such as CNN (LeCun *et al.*, 1998; Krizhevsky *et al.*, 2012; Simonyan, 2014), GAN (Goodfellow *et al.*, 2020; Radford *et al.*, 2015; Zhu *et al.*, 2017), DDPM (Ho *et al.*, 2020; Song *et al.*, 2020; Nichol and Dhariwal, 2021), and GAI (Vaswani, 2017; Devlin, 2018; Brown, 2020), enhance the adaptability and accuracy of digital twins by efficiently handling complex datasets and uncovering insights beyond the capabilities of traditional models. These converging advancements have transformed digital twins from a visionary concept into a practical, scalable tool, supporting a proactive, patient-centered approach that significantly elevates the precision and personalization of modern healthcare.

## 1.1 Importance and Advantages of This Review

As digital twin technology advances rapidly and its applications in healthcare broaden, it becomes increasingly important to understand its current development as shown in Figure 2. This review aims to meet that need by examining the system-specific applications of digital twins and highlighting their potential to revolutionize healthcare through personalized medicine. Our primary focus is on the latest computational models used in creating digital twins, providing a detailed exploration of the methods for constructing these twins and refining the associated algorithms and model architectures. Our analysis covers the entire lifecycle—from data collection and simulation to predictive analytics—offering a comprehensive yet accessible overview for both computer scientists seeking technical depth and medical professionals interested in clinical applications.

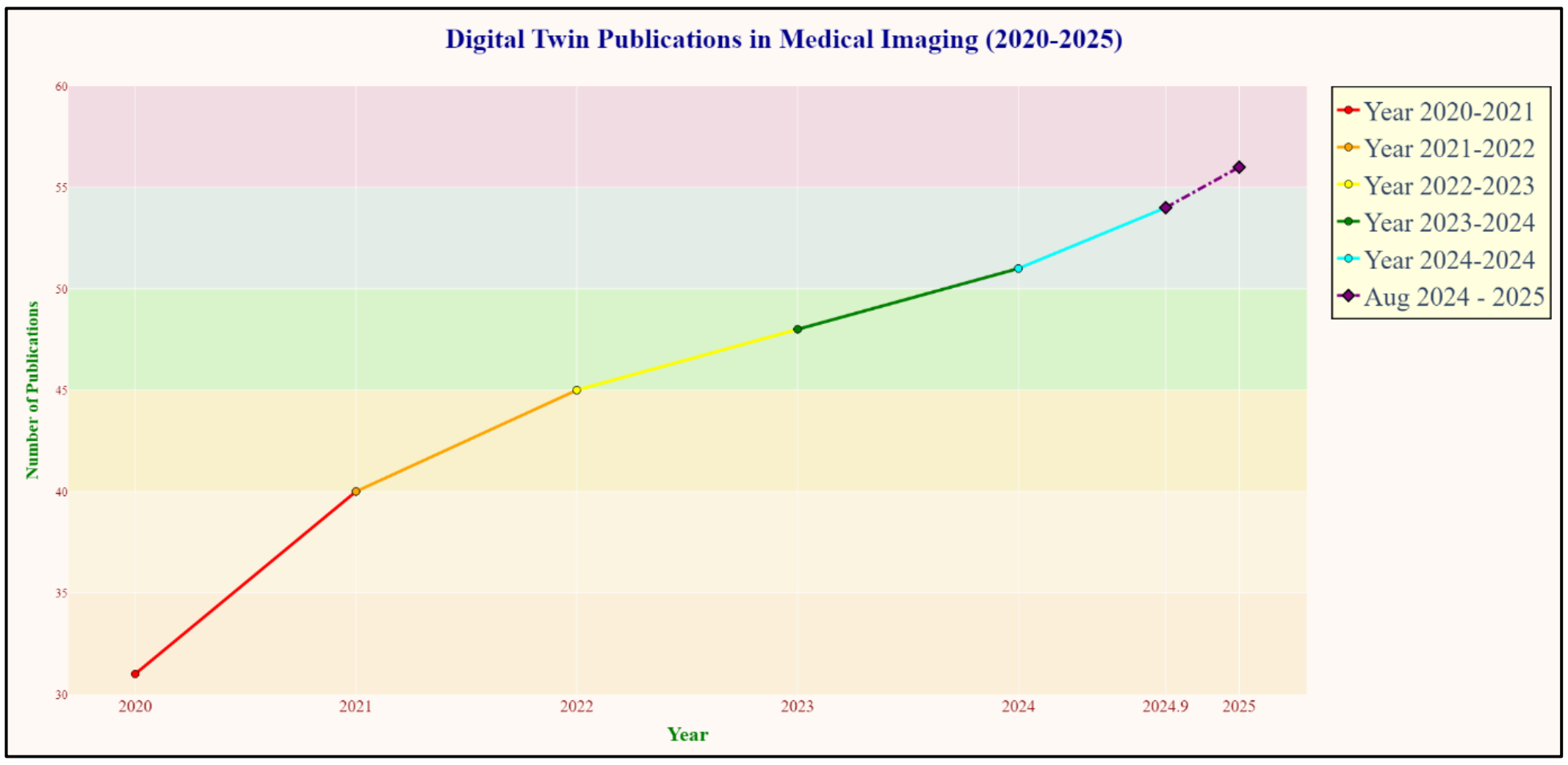

**Figure 2**. Steady growth in research focus on digital twin technology in medical imaging.

Central to this paper is a comprehensive analysis of the computational modeling techniques identified in the selected body of research. These techniques encompass image processing algorithms tailored to extract critical features from medical images, time-series analysis for detecting temporal patterns within patient health data, and advanced simulation methodologies such as FEA and CFD. These sophisticated tools substantially enhance the precision and functionality of digital twins, facilitating real-time monitoring and predictive capabilities. By incorporating these state-of-the-art approaches, this review offers a rigorous and analytically sound examination that highlights the transformative potential of digital twins in the realm of personalized medicine.

One particularly novel aspect of our review is the focus on digital-twin human system development based on medical imaging, a perspective not commonly addressed in current literature. Creating a digital twin of the entire human body is a complex and challenging endeavor, comparable to assembling an intricate puzzle. To tackle this challenge, researchers have shifted toward focusing on system-specific digital twins, a more practical and manageable approach. We examine studies where researchers have broken the body down into smaller, more manageable components, developing detailed, actionable models for individual organs or systems. This focused approach allows for a level of precision and insight that broader, whole-body models cannot achieve. Each system-specific twin serves as a specialist, tailored to address the unique challenges of its respective system—whether in heart function, liver health, or other processes—contributing to a more comprehensive understanding of human health.

In conclusion, through detailed examples and thorough exploration, we demonstrate the unique strengths of our review. We aim not only to inform but also to inspire, encouraging further research and innovation in this groundbreaking field. Positioned at the intersection of technology and medicine, our review seeks to shape the future of digital twin technology in healthcare, particularly in the realm of medical imaging. By providing insights that are both technically rigorous and practically relevant, we hope to pave the way for advancements that will improve patient outcomes and redefine the landscape of personalized medicine.

## 2. System-Specific Review

In this section, we explore the system-specific aspects of our review, providing an in-depth analysis of various bodily systems. By focusing on the construction of digital twins using medical imaging data and the advancements in the algorithms and model architectures that support and refine them for higher precision, we aim to present a detailed understanding of the latest developments and emerging research trends in this application. It offers valuable insights for researchers and healthcare practitioners working to advance the field of personalized medicine.

### 2.1 Methodology for Literature Review

To comprehensively gather relevant literature, we conducted a PRISMA (Page *et al.*, 2021) protocol to systematic search the peer-reviewed journal articles across Scopus, PubMed, Web of Science, and Google Scholar, using targeted queries that combined terms such as "Digital Twin" and "Digital Replica" with medical imaging modalities, including MRI and CT. This approach yielded 117 records from Scopus, 115 from PubMed, and 17 from Web of Science. Google Scholar search initially returned over 800 records but only 80 were deemed highly relevant. After removing duplicates, we obtained a total of 260 peer-reviewed journal articles as initial identification as shown in Figure 3. After initial screening, we subsequently assessed 181 full-text articles for eligibility, and excluded 97 articles for various reasons: 3 were not in English, 6 focused solely on ethical concerns, 1 was pre-clinical, 15 were unrelated to medical imaging, and 87 lacked sufficient technical detail. Ultimately, 69 studies met our inclusion criteria and were incorporated into the qualitative synthesis.

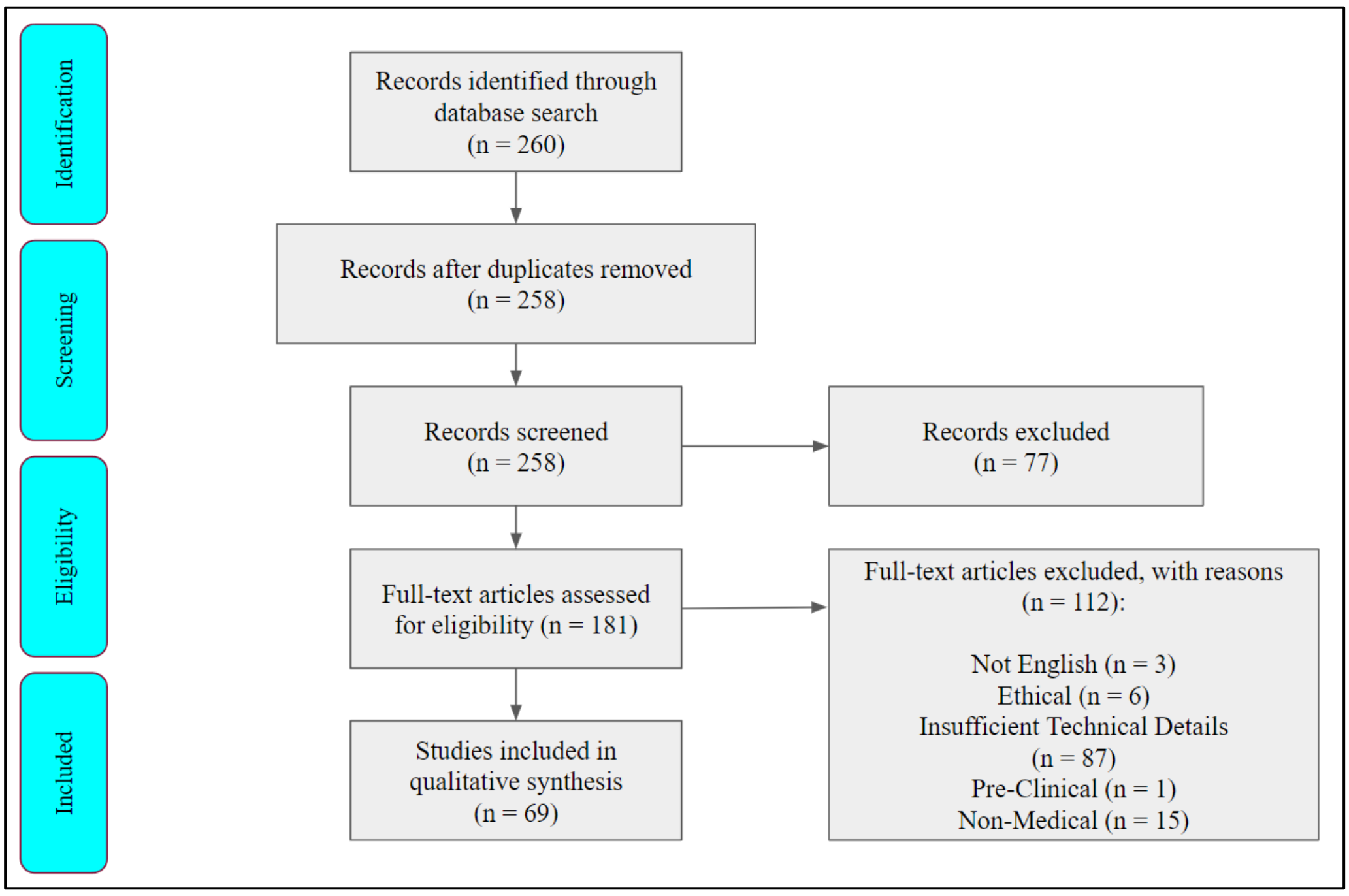

**Figure 3**. PRISMA flow diagram.

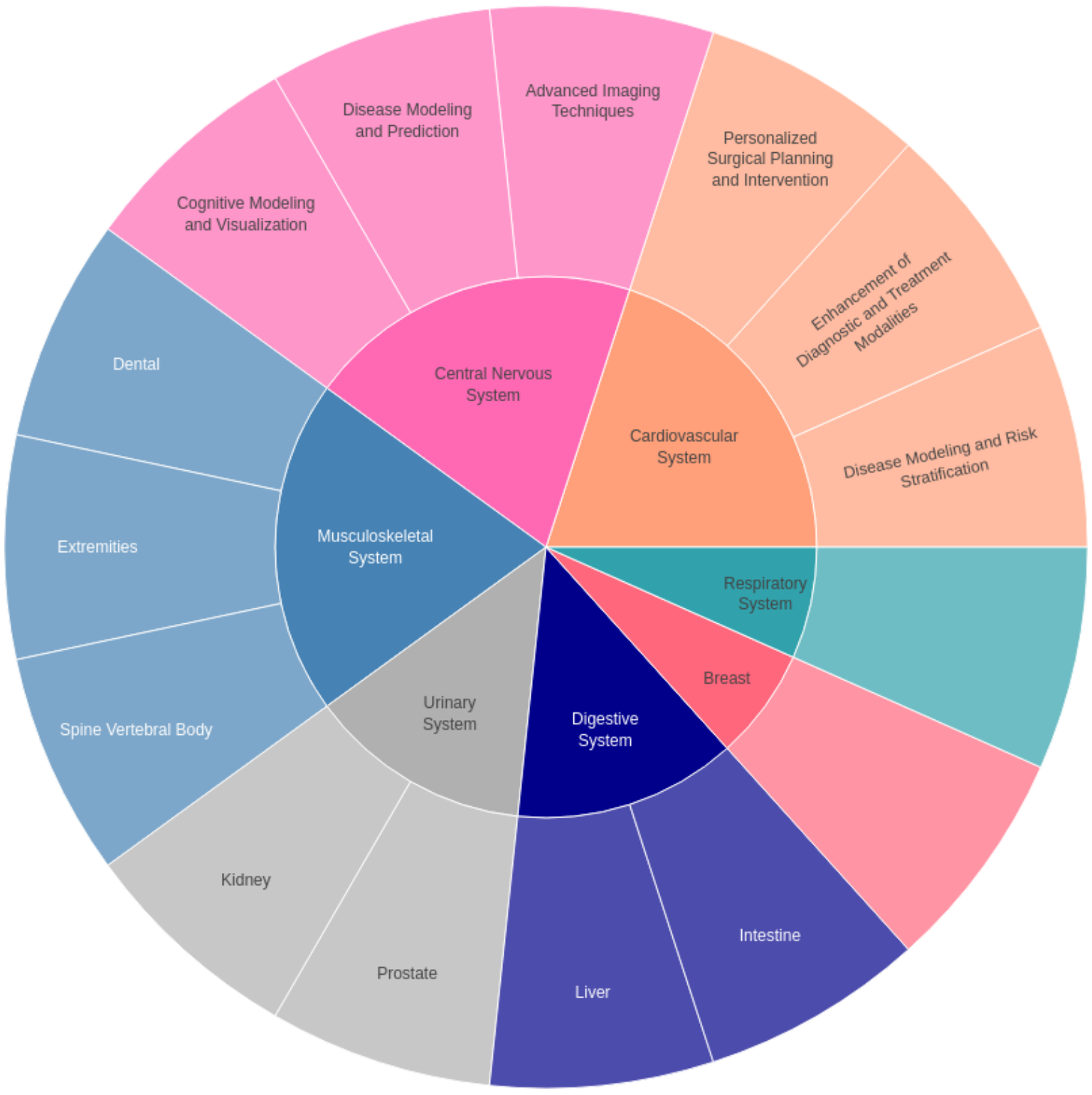

**Figure 4**. Classification of digital twin applications across medical fields.

Based on the qualified research work from Figure 3, we further classified various digital-twin systems into Figure 4, which visually represents the diverse applications in medical fields. Each segment corresponds to a specific medical domain, such as cardiovascular, respiratory, breast, digestive, liver, musculoskeletal, and others. Within each segment, subdivisions outline particular application scenarios of digital twin technology within that field, including diagnostic enhancement, personalized treatment planning, image processing, simulation, cognitive modeling, and more. The chart provides a comprehensive visual overview of the classification methodology, offering insight into the extensive applications and potential research trajectories for digital twin technology in healthcare. This paper is structured around a classification analysis of different systems within the human body, which forms the foundation of this system-specific review. By adopting this classification framework, we systematically examine the applications and advancements of digital twin technology across various medical domains, allowing for a focused and detailed exploration of its impact and potential in healthcare.

## 2.2 Cardiovascular System

In cardiovascular digital twin research, MRI and CT imaging are predominantly used as primary data sources, especially when investigating arterial stenosis and various cardiac diseases. Occasionally, additional imaging modalities like CTA, DUS, and X-ray are incorporated to provide supplementary diagnostic information. The majority of datasets utilized in these studies are not publicly accessible, limiting widespread availability; however, certain initiatives are working toward sharing specific datasets. From a computational perspective, there is a strong dependence on conventional modeling techniques such as CFD and FEA for simulating hemodynamic processes. A significant shift toward employing advanced machine learning models—like deep neural networks and generative algorithms—is evident to more effectively capture complex physiological interactions. Furthermore, the combination of mechanistic and data-driven modeling approaches is increasingly adopted to enhance the accuracy and adaptability of cardiovascular digital twins.

**Table 2.** Summary of digital-twin studies for cardiovascular system.

| Reference | Clinical Site | Data Modality | Data Quantity | Data Publicity | Computational Models |
|---|---|---|---|---|---|
| (Banerjee *et al.*, 2021) | Myocardial Infarction | Cine MRI | 20 Patients | Not public | FCN, SSM |
| (Dubs *et al.*, 2023) | Carotid Artery Stenosis | CTA, DUS | 26 Patients, 37 Bifurcations | Available Upon Request | CFD |
| (Loewe *et al.*, 2022) | Arrhythmias, HF | MRI, CT, Electrophysiology | N/A | Not Public | CFD, EP |
| (Rouhollahi *et al.*, 2023) | Aortic Stenosis | CT | 35 Patients | Not Public | U-Net, DCNN |
| (Renaudin *et al.*, 1994) | Coronary Artery Stenosis | MRI, X-ray | 24 Phantoms | N/A | Computer-Aided Design, CAM, Stereolithography |
| (Galappaththige *et al.*, 2022) | Cardiac EP | MRI, CT | 24 Patients | Available On Zenodo | Cardiac EP, VVUQ |
| (Liang *et al.*, 2022) | DORV | CT, Echocardiography | 6 Patients | Not Public | Virtual Models |
| (de Lepper *et al.*, 2022) | Ischaemic Cardiomyopathy | ECG, MRI, CT | N/A | Not Public | Data-Driven Models, Mechanistic Models |
| (Straughan *et al.*, 2023) | Coronary Artery Disease | OCT | 28 Patients | Not Public | FEA, CNN, 3D Reconstruction |
| (Rudnicka *et al.*, 2024) | AF | MRI, CT, ECG | N/A | Not Public | DNN, YOLO, GANs, GNNs |
| (Vanzella *et al.*, 2024) | Myocardial Infarction, Hypertension, Dyslipidemia | N/A | 23 Patients | Not Public | Thematic Analysis |
| (Chakshu *et al.*, 2021) | AAA | MRI, CT, Blood Pressure Waveforms | 4137 Virtual Patients | Not Public | LSTM, CNN, Inverse Analysis |
| (Trayanova and Prakosa, 2024) | AF, V, SCD | MRI, ECG | N/A | Not Public | ML models, Mechanistic models |
| (Bodin *et al.*, 2021) | Coronary Artery Disease, Myocardial Ischemia | ECG, 3D Imaging | N/A | Not Public | Aliev-Panfilov Model, Voxel Models, 3D Reconstruction |
| (Koopsen *et al.*, 2024) | Left Bundle Branch Block, Myocardial Infarction | MRI, Echocardiography | 13 Patients | GitHub Repository Available | CircAdapt Model, Particle Swarm Optimization |
| (Pires *et al.*, 2024) | Ebstein Anomaly | 3D Ultrasound | 1 Patient | Not Public | 3D Virtual, 3D Printing |
| (Ložek *et al.*, 2024) | Tetralogy of Fallot | Echocardiograph, MRI, CT | 2 Patients | Available Upon Request | CircAdapt |

### 2.2.1 Cardiovascular Disease Modeling and Risk Stratification

In recent years, the integration of advanced technologies has catalyzed significant improvements in cardiovascular healthcare.

Advanced modeling techniques, combined with digital twin technology, offer substantial advancements in Computer-Aided Design modeling, enabling precise risk stratification and supporting personalized clinical decision-making, ultimately contributing to better patient outcomes. The study by Renaudin et al. (Renaudin *et al.*, 1994) exemplifies this transformation through a novel approach involving stereolithography-based Computer-Aided Design/CAM to create highly precise three-dimensional arterial phantoms. These phantoms, characterized by sub-0.1 mm accuracy, provide a critical foundation for accurately modeling coronary arteries, facilitating the integration of these detailed models into digital twin frameworks.

Straughan et al. (Straughan *et al.*, 2023) extended these efforts by automating the construction of finite element simulations using OCT data. This innovation allows for efficient and precise stress analysis of coronary lesions, showcasing the potential to transform patient-specific imaging into actionable digital twin models for CAD assessment. Similarly, Chakshu et al. (Chakshu *et al.*, 2021) employed LSTM-based inverse analysis to derive blood pressure waveforms non-invasively, effectively integrating deep learning with digital twins for real-time cardiovascular monitoring. The contributions by Koopsen et al. (Koopsen *et al.*, 2024) further demonstrate the power of digital twins by creating personalized models of left ventricular mechanical discoordination, enhancing the reliability of these models through parameter subset reduction and identifiability analysis.

Further evidence of the transformative power of digital twin technology in heart disease modeling and treatment is provided by studies in related cardiovascular domains. Loewe et al. (Loewe *et al.*, 2022) developed a highly personalized computational heart model using CT and MRI data to simulate electrical activity and assess SCD risk, demonstrating a hazard ratio significantly higher than that of traditional clinical metrics. This underscores the superiority of digital twins in heart failure modeling and precise risk stratification. Moreover, Banerjee et al. (Banerjee *et al.*, 2021) applied deep learning techniques to 2D cine MR images for arrhythmia treatment, generating accurate 3D heart models by correcting misalignments, which enhanced the precision of electrophysiological simulations and diagnosis. Additionally, de Lepper et al. (de Lepper *et al.*, 2022) introduced a hybrid modeling approach that combined mechanistic and data-driven models, enabling better identification of high-risk VT patients compared to traditional metrics, thereby reducing unnecessary ICD implantation.

### 2.2.2 Enhancement of Diagnostic and Treatment Modalities

Recent studies demonstrate how digital twin technology is revolutionizing diagnostic imaging, making substantial contributions to both clinical research and practical applications. Bodin et al. (Bodin *et al.*, 2021) exemplify this integration by demonstrating the use of digital twins through the transformation of ECG data into visualizations of electrical activity within a 3D heart model. This innovative approach not only enabled precise simulation of myocardial damage but also provided a tangible framework for understanding the spatial dynamics of cardiac conditions, thereby helps clinicians better identify pathological regions and directly enhancing diagnostic accuracy and informing clinical decision-making.

Rouhollahi et al. (Rouhollahi *et al.*, 2023) built upon this foundation, highlighting the role of digital twins in cardiovascular diagnostics through a fully automated deep learning tool, CardioVision. Utilizing CT images, CardioVision effectively generated precise 3D reconstructions of the aortic valve, achieving highly accurate calcification quantification. This digital twin model significantly reduced manual effort and expedited the analysis process, demonstrating the feasibility of large-scale clinical applications. The ability to automate image segmentation and produce detailed 3D reconstructions is pivotal in advancing the scalability of cardiovascular diagnostics, particularly in complex cases such as aortic stenosis.

Pires et al. (Pires *et al.*, 2024) extended these capabilities further into the realm of congenital heart disease by leveraging obstetrical ultrasound data to create 3D digital twins of fetal hearts. These digital models provided an exceptionally detailed visualization of congenital anomalies, such as Ebstein anomaly, which played a crucial role in aiding pre-surgical planning and family counseling. The ability to faithfully replicate the intricate anatomical structures of fetal hearts enabled medical teams to observe and assess abnormalities with a level of clarity that traditional 2D ultrasound could not provide. Such detailed digital twins bridge the gap between diagnostic imaging and actionable clinical insights, underscoring their value in personalized medicine.

In addition to surgical planning, the role of digital twin technology extends to non-invasive diagnostics and validation. Dubs et al. (Dubs *et al.*, 2023) leveraged CFD to construct patient-specific digital twins of carotid bifurcations, enabling a more precise, non-invasive assessment of CAD severity. The CFD-based digital twin models offered detailed insights into hemodynamic parameters, such as pressure drops across stenoses under varying flow conditions, which significantly enhanced diagnostic accuracy compared to traditional diagnostic tools. Meanwhile, Galappaththige et al. (Galappaththige *et al.*, 2022) expanded on the application of digital twins by developing personalized ventricular models for cardiac electrophysiological applications. By focusing on patient-specific differences and employing VVUQ methods, the study emphasized the necessity of personalized modeling for ensuring the credibility and applicability of cardiac simulations across different patient profiles. Collectively, these studies demonstrate how digital twin technology enables the creation of highly individualized cardiac models and provides a robust framework for validating cardiovascular interventions.

### 2.2.3 Personalized Surgical Planning and Intervention

Emerging studies demonstrate that digital twin technology is reshaping personalized treatment plans, surgical planning, and rehabilitation, contributing to remarkable advancements in cardiovascular disease management. Trayanova and Prakosa (Trayanova and Prakosa, 2024) exemplified this transformation by utilizing multimodal data—including ECG, CT, and MRI—to create personalized digital twin models of the heart. These models significantly enhanced the precision of arrhythmia treatment by accurately predicting optimal ablation targets, leading to improved outcomes for patients with atrial fibrillation. This study highlights the power of digital twins to bridge the gap between diagnostic imaging and actionable clinical interventions. Similarly, Ložek et al. (Ložek *et al.*, 2024) applied digital twin technology to model right ventricular dyssynchrony in patients post-Tetralogy of Fallot repair. Their findings demonstrated that using digital twin simulations to personalize CRT led to notable improvements in ventricular function and efficiency, offering compelling evidence in favor of data-driven, individualized therapeutic decisions.

In addition to treatment planning, digital twin technology has demonstrated significant value in surgical planning and rehabilitation, thereby expanding its influence across various stages of cardiovascular disease management. Rudnicka et al. (Rudnicka *et al.*, 2024) illustrated the application of digital twins in surgical planning by leveraging AI and XR technologies to model cardiac functions. This integration enhanced diagnostic precision, optimized targeted therapy selection, and reduced uncertainty in surgical procedures, thereby improving the reliability of treatment outcomes. Meanwhile, Vanzella et al. (Vanzella *et al.*, 2024) highlighted the use of digital twins in cardiac rehabilitation, employing hybrid and virtual models to address the unique challenges posed by the COVID-19 pandemic. These models not only improved patient adherence but also enhanced quality of life by overcoming barriers such as accessibility and convenience.

Recent studies have shown that advancements in digital twin technology significantly enhance the accuracy and dependability of surgical planning, testing, and validation in cardiovascular disease management. Liang et al. (Liang *et al.*, 2022) introduced an innovative approach combining virtual models with 3D printing to simulate the DORV surgical procedure, resulting in a substantial reduction in diagnostic uncertainties and errors. By employing these digital twins as auxiliary tools, the study achieved a significant decrease in the rate of uncertain or incorrect diagnoses from 42.5% to 4.6%. This finding illustrates the transformative capacity of digital twins to offer a clear and tangible representation of complex cardiac anatomies, ultimately guiding surgical interventions with greater accuracy and confidence.

### 2.3 Central Nervous Systems

Among the 12 studies on central nervous system digital twins, as seen in Table 3, MRI is the primary modality, essential for capturing brain anatomy and activity (Glover, 2011), often combined with other techniques like DTI and fMRI. With abundant datasets available for prevalent diseases like Alzheimer's disease, tumors, and multiple sclerosis, researchers can explore these topics further with rich resources, in contrast to whole-brain simulations highlight a need for more publicly available data. The computational models range from traditional methods like K-Means for clustering to modern approaches such as NeuralODEs and Spiking Neuronal Networks for dynamic modeling, with model complexity tailored to specific clinical needs.

**Table 3.** Summary of digital-twin studies for central nervous systems.

| Reference | Clinical Site | Data Modality | Data Quantity | Data Publicity | Computational Models |
|---|---|---|---|---|---|
| (Cen *et al.*, 2023) | Multiple sclerosis | MRI | 2572 Patients | Available Upon Request | Mixed Splines |
| (Hu *et al.*, 2021) | Tumor - General | MRI | N/A | Available Upon Request | HPU-Net |
| (Wan *et al.*, 2021) | Tumor - General | MRI | N/A | Available Upon Request | SVM + AlexNet |
| (Sarris *et al.*, 2023) | Tumor - General | MRI | 247 Patients | Available at Radiopaedia and Kaggle | K-Means |
| (Wang *et al.*, 2024) | Tumor - General | Optical Imaging (MCF Endoscopy), Autofluorescence Imaging | 50 Patients | Available Upon Request | U-Net, EDSR Network, Transfer Learning |
| (Guillevin *et al.*, 2023) | Tumor - Gliomas | MRI + DTI | N/A | N/A | Graph Theory |
| (Chaudhuri *et al.*, 2023) | Tumor - Gliomas | MRI | 100 In Silico Patients | Available Upon Request | Bayesian Model |
| (Wang *et al.*, 2023) | Whole brain simulation | MRI | N/A | N/A | CNNs + GAN |
| (Lu *et al.*, 2023) | Whole brain | MRI + DTI + PET | 1 Patient | Not public | Spiking Neuronal Networks |
| (Lu *et al.*, 2022) | Whole brain | MRI | 1 Patient | Not public | Hierarchical Mesoscale Data Assimilation |
| (Li *et al.*, 2023) | Whole brain | MRI + fMRI | N/A | N/A | Hierarchical Mesoscopic Data Assimilation |
| (Lachinov *et al.*, 2023) | Alzheimer's disease | MRI + OCT | 876 Patients | Available partially at ADNI and TADPOLE Dataset | NeuralODEs |

The CNS, encompassing the brain and spinal cord, is one of the most complex and delicate systems in the human body. Accurately simulating and representing CNS functions is therefore essential for advancing our understanding of neurological diseases, improving diagnostic capabilities, and enabling personalized treatment strategies. Some, like Wan et al. (Wan *et al.*, 2021), utilize digital twin technology with semi-supervised learning model to enhance brain image fusion and diagnostic accuracy, enabling digital twins to perform feature recognition and forecasting tasks by mapping real-time brain imaging data into virtual space. In addition to diagnostic improvement, people are exploring the use of digital twins to support precise diagnostics, optimize personalized treatment plans, enhance medical imaging, facilitate cognitive simulations, and enhance visualization and neurosurgical applications, revolutionizing CNS healthcare.

### 2.3.1 Advanced Imaging Techniques for Digital Twin Models

Just as a human doctor relies on fine-detailed images for accurate diagnosis, digital twins also benefit from high-resolution imaging to capture anatomical information for precise modeling. With this objective, Wang and Qiao (Wang *et al.*, 2023) conducted a novel approach to enhance MRI quality using several super-resolution techniques, including Deep Transfer Learning and GANs. Their model, which incorporates an adaptive decomposition fusion algorithm, improved MRI resolution and detail significantly. Additionally, they utilized metamaterial technology to further boost MRI signal quality, achieving metrics such as a PSNR of 34.11 dB and an SSIM of 85.24%, primarily on brain MRI images. This study suggests broader applications in digital twin models, offering enhanced spatial and spectral fidelity for improved diagnostics.

Similarly, Wang and Dremel (Wang *et al.*, 2024) explored a deep learning approach to enhance tumor imaging resolution with MCF endoscopes. Using a cascaded network architecture with U-Net and an EDSR Network, they pre-trained the model on simulated MCF images and fine-tuned it on glioblastoma tissue data, achieving PSNR values up to 31.6 dB and SSIM of 0.97. This method contributes to digital twin development by providing higher-resolution visualizations crucial for tumor analysis.

Building on these refined images, digital twin models can now be employed for a variety of applications, including precise disease modeling, progression prediction, and cognitive simulation.

### 2.3.2 Disease Modeling and Prediction

One of the purposes of having a digital twin for the central nervous system is to detect the disease, model its progression, and make predictions for proactive treatment planning. In 2023, Sarris et al. (Sarris *et al.*, 2023) investigated the use of K-means clustering within digital twin technology to enhance brain tumor detection. A 3D model of the brain was constructed using open-source computer-aided design, where tumor tissues were differentiated from normal brain tissues through clustering of pixel intensity variations. Similarly, Cen and colleagues (Cen *et al.*, 2023) tackled the challenge of predicting disease-specific brain atrophy in MS by employing digital twin technology. Due to the scarcity of extensive longitudinal brain MRI data, the study utilized a spline model to create normal aging trajectories as a baseline for comparison against MS patient data, allowing for the prediction of thalamic atrophy onset up to 5-6 years before clinical symptoms appear.

Building on these detection techniques, Hu et al. (Hu *et al.*, 2021) introduce an advanced fuzzy system within digital twin technology to enhance brain MRI diagnostics, particularly for brain tumors. Unlike traditional binary logic systems, fuzzy systems accommodate more nuanced values, mimicking human reasoning and improving predictive accuracy. They combined this enhanced fuzzy clustering with HPU-Net for MRI segmentation, achieving high metrics such as a DSC of 0.936 and a Jaccard coefficient of 0.845. Powered by CNN and RNN, their system offers improved segmentation accuracy and significantly reduces noise. These advanced techniques help digital twin models better handle real-world complexity, laying a strong foundation for predictive disease modeling.

Following disease detection and diagnosis, ongoing monitoring and prediction become essential. Wan and Dong (Wan *et al.*, 2021) applied semi-supervised learning with digital twin technology to enhance brain tumor analysis. By combining S3VM and an enhanced AlexNet model, the study achieved 92.52% accuracy in tumor segmentation and feature extraction from MRI data.

After predicting the presence of tumors, tracking their progression is highly valuable. For instance, Guillevin et al. (Guillevin *et al.*, 2023) used multimodal MRI, including fMRI, DTI, and MRS, to develop a digital replica of gliomas. These models simulate glioma progression and therapeutic response, aiding in personalized treatment planning. Chaudhuri et al. (Chaudhuri *et al.*, 2023) also presented a progression-predictive digital twin framework for high-grade gliomas, dynamically incorporating updated patient MRI data with a Bayesian model to predict tumor growth. This framework not only forecasts tumor progression but also adapts radiotherapy regimens based on patient-specific data, accounting for uncertainties in tumor biology. With digital twins, we can not only detect brain diseases such as tumors early on but also model their progression over time, enabling continuous monitoring and personalized treatment planning.

### 2.3.3 Cognitive Modeling and Visualization

Apart from neoplastic diseases like tumors, digital twins are also valuable for studying neurodegenerative conditions. A solid understanding of brain and disease simulations is enhanced by visualizing the brain's digital twin, making it more practical for clinical and research applications. To simulate human brain function, Lu et al. (Lu *et al.*, 2022) developed a comprehensive DTB model, starting with their 2022 paper, which established a framework to replicate brain states and perform tasks like virtual deep brain stimulation. This model incorporates mesoscale data assimilation to simulate both resting and active states, laying the groundwork for cognitive task modeling. In 2023, Lu et al. (Lu *et al.*, 2023) expanded on this foundation by improving computational efficiency and applying the DTB to real-world cognitive experiments, demonstrating its potential for advanced research and clinical applications. Capable of simulating 86 billion neurons, this model requires substantial computational resources: over 3,500 computer nodes and more than 10,000 GPUs.

Similarly, Li et al. (Li *et al.*, 2023) introduced the DTBVis system, an interactive platform for comparing human and digital twin brains. DTBVis includes four visualization modules: the main 3D view for topological connectivity and BOLD signals, the auxiliary view for FC matrices or t-SNE plots, the detailed view for in-depth BOLD signals, and the timeline view for tracking temporal changes. This system allows users to explore individual brains or directly compare two, highlighting differences in BOLD signals to assess the DTB's accuracy against real human data. By advancing simulation and cognitive modeling, DTBVis is a valuable tool in digital twin and medical imaging research within neuroscience.

Being able to simulate general brain functions, digital twin models are applied to neurodegenerative diseases. AD, a significant public health concern, affects millions worldwide (Knopman *et al.*, 2021). The paper by Lachinov et al. (Lachinov *et al.*, 2023) addresses disease progression in AD, using MRI to monitor anatomical changes. Lachinov et al. utilized CNNs to extract features from MRI images and then applied NeuralODEs to predict the continuous evolution of disease-related anatomical changes. Their framework leverages longitudinal data to simulate dynamic progression, offering a robust model for neurodegenerative diseases. Notably, their model extends beyond AD, applying to other conditions such as GA, where OCT captures retinal degeneration. This versatility underscores the digital twin model's potential for broad applications in predicting the progression of various degenerative diseases.

### 2.4 Musculoskeletal System

Many studies rely on CT or its variants (e.g., Cone Beam CT, Micro-CT) for digital twin construction in bones, joints, and musculoskeletal structures, as shown in Table 4, due to CT's superior capability for capturing detailed anatomy (Engelke *et al.*, 2018). Other modalities like Optical Imaging and EMG are used selectively, but CT remains the primary choice. There are some dataset available for extremities and spine studies, but there is limited availability for dental applications. For computational models, dental studies favor traditional models like FEA, extremities use mixed statistical and 3D modeling approaches, and spine studies increasingly apply neural networks and generative models.

**Table 4.** Summary of digital-twin studies for musculoskeletal system.

| Reference | Clinical Site | Data Modality | Data Quantity | Data Publicity | Computational Models |
|---|---|---|---|---|---|
| (Ahn *et al.*, 2024) | Dental - Implants and Surrounding Bone | Cone Beam CT (CBCT) | 25 Patients | Not public | Parametric ROM |
| (Demir *et al.*, 2023) | Dental - Mandible | CT | 13 Patients | Not public | Non-linear FEA |
| (Kim *et al.*, 2018) | Dental - Teeth | 3D dental scans | N/A | Not public | Statistical analysis - Spearman's Correlation Analysis |
| (Vila *et al.*, 2024) | Extremities - Bone allografts | CT | N/A | Not public | 3D modeling based on DICOM standards |
| (Ioussoufovitch and Diop, 2024) | Extremities - Finger joints | Optical Imaging | N/A | Available upon request from authors | Spatiotemporal Fourier Component Analysis, Monte Carlo Simulations |
| (Hernigou *et al.*, 2021) | Extremities - Foot and Ankle | CT | 5 Patients | Available upon request from hospital | CAD Catia™ |
| (Kim *et al.*, 2023) | Extremities - Shoulder | CT | N/A | Not public | Shapekey Function in Blender, Unity Engine |
| (Verweij *et al.*, 2024) | Extremities - Shoulder | CT | 24 Patients | Not public | 3D Modeling and Segmentation |
| (Ha *et al.*, 2024) | Extremities - Upper Arm | EMG, Thermography | N/A | Available | SVM |
| (Moztarzadeh *et al.*, 2023) | Spine Vertebral Body - Cervical Vertebrae | X-ray | 319 Patients | Available upon request from authors | MobileNetV2 |
| (He *et al.*, 2021) | Spine Vertebral Body - Lumbar Spine (L2-L5) | CT | N/A | Available | FEM, GPR |
| (Ahmadian *et al.*, 2022a) | Spine Vertebral Body - Vertebra (Spine) | CT | 1 Patient | Not public | ReconGAN (DCGAN with Finite Element Analysis) |
| (Ahmadian *et al.*, 2022b) | Spine Vertebral Body - Vertebra (Spine) | CT, Micro-CT | N/A | Not public | ReconGAN (DCGAN with Finite Element Analysis) |
| (Swaity *et al.*, 2024) | Dental - Maxillary Canine | CBCT | 100 Patients | Available upon request from authors | 3D UNet |

The musculoskeletal system, encompassing bones, muscles, tendons, and ligaments, is essential for movement, stability, and support. Its health is crucial for physical function and quality of life. Digital twin technology offers significant benefits for the musculoskeletal system by enabling precise, personalized models for diagnosing conditions, planning surgeries, and monitoring recovery. Same as other systems, medical imaging, such as CT, MRI, and 3D scanning, plays a key role in constructing these digital twins, allowing accurate visualization and simulation of musculoskeletal structures and their biomechanical behaviors. In the following sections, we will explore current digital twin applications in Dentistry, extremities, and Spine Vertebral Body.

### 2.4.1 Dental

In dentistry, digital twin technology has undergone a transformative evolution, with pioneering research extending its applications from basic clinical practices to complex personalized treatments. It commenced with the foundational work by Zilberman et al. (Zilberman *et al.*, 2003), which validated the accuracy and clinical relevance of virtual orthodontic models by comparing measurements from digital twins against traditional caliper measurements on plaster models. This initial validation set the stage for more intricate applications in clinics. Such applications address conditions like periodontal disease, which affects 20-50% of the global population (Nazir, 2017).

Building on these initial validations, subsequent research has focused on digital twins in dentistry focusing on enhancing diagnostic and surgical precision. For instance, Kim et al. (Kim *et al.*, 2018) used the Opto TOP-HE 3D scanner to create detailed virtual models for measuring tooth wear. By repeating scans 10-20 times, they generated full-mouth dental casts, which were then rendered and analyzed with Rapidform software. In contrast to the statistical methods used by them, which involved repeated scanning, the study by Swaitly et al. (Swaity *et al.*, 2024) represents a significant advancement in dental imaging through deep learning. Their approach utilizes a U-Net architecture to automate the segmentation of maxillary impacted canines directly from CBCT images. The model achieved a DSC of 0.99, which is nearly perfect between inference and ground truth. This method enhances the precision of the 3D models generated and drastically reduces the time required for segmentation. Swaitly asserts that the creation of fast and reliable 3D models supports the implementation of digital twins in personalized dental care.

Beyond orthodontics, digital twins have expanded their applications to implant dentistry, where the recent study by Ahn et al. (Ahn *et al.*, 2024) introduces a novel parametric ROM that leverages finite element analysis to simulate real-time stress distributions in dental implants and surrounding bone structures. By integrating patient-specific medical imaging data from CBCT scans into a digital twin framework, this model enables dentists to visualize and modify implant placement plans dynamically, enhancing the precision and success of implant surgeries.

In the mandible, Demir et al. (Demir *et al.*, 2023) used CT scans to create a digital twin of the human lower jaw, employing the Mimics Innovation Suite to segment cortical and cancellous bone based on HU thresholds. They optimized the mesh for efficiency and anatomical accuracy, and then assigned material properties, such as elasticity modulus and Poisson's ratio, to simulate biomechanical behavior accurately. This approach allowed the digital twin to predict the mandible's biomechanical response under impact conditions, enhancing its clinical relevance for maxillofacial trauma analysis.

### 2.4.2 Extremities

Extremities disorders, such as arthritis and fractures, significantly impact mobility and quality of life, with arthritis affecting 21.2% of adults in the United States (Fallon, 2023). Digital twin technology is transforming the management of these conditions by providing clinicians with precise visualization and simulation tools to better understand and treat complex musculoskeletal issues. This technology enables detailed anatomical insights, enhances surgical precision, optimizes allograft matching, and supports real-time rehabilitation and monitoring. The following subsections explore four critical areas where digital twins are advancing limb healthcare: Visualization and Simulation, Precision in Surgical Interventions, Allograft Matching, and Real-Time Rehabilitation and Monitoring.

Visualization and simulation are crucial for understanding musculoskeletal anatomy, providing clinicians with dynamic tools to manage complex conditions effectively. In shoulder treatment, Kim et al. (Kim *et al.*, 2023) created interactive 3D visualizations of shoulder mechanics, using CT scans converted to STL format for detailed modeling in Blender. Through animation, the models simulated shoulder movements and were integrated into Unity, enabling an interactive application that provides comprehensive insights into shoulder muscles such as the Deltoid, Infraspinatus, and Supraspinatus. This detailed simulation aids clinicians and patients in visualizing shoulder mechanics for more effective management and rehabilitation. Similarly, Verweij et al. (Verweij *et al.*, 2024) developed virtual models of the humerus and scapula to analyze shoulder instability after anterior dislocations. Using 3D segmentation and registration, they positioned the humeral head within a glenoid coordinate system, enabling precise measurements of joint space and positioning.

Building on the anatomical insights gained from visualization and simulation, digital twins are instrumental in enhancing surgical precision, particularly in lower limb procedures like total ankle arthroplasty and knee surgery. Hernigou et al. (Hernigou *et al.*, 2021) employed CT scans to refine implant placement, converting CT images to DICOM format before creating detailed 3D models using CAD tools like Catia™. This digitalization process seamlessly integrates high-fidelity anatomical data into the digital twin environment, where artificial intelligence and machine learning analyze and simulate kinematic behaviors. Such precision in modeling and simulation optimized implant placement and contributing to personalized orthopedic care.

After precise surgical interventions, ensuring accurate anatomical compatibility in bone grafting is essential for successful outcomes. Vila et al. (Vila *et al.*, 2024) introduced the BeST-Graft Viewer, a digital twin-based tool that optimizes the matching of bone allografts to recipient sites, specifically in femoral applications for lower-limb surgeries. This system digitizes grafts through CT scans, storing data in DICOM format, and reconstructs 3D models for detailed anatomical assessments in 2D, 3D, and multiplanar views. The BeST-Graft Viewer automates the graft-recipient matching protocol, facilitating precise measurements and minimizing manual errors. By integrating real-time data management with interactive visualization, this system exemplifies how digital twins enhance orthopedic care, where anatomical matching precision is crucial to the success of grafting procedures.

After surgery and precise allograft matching, rehabilitation and continuous monitoring are crucial for ensuring successful recovery and long-term functional improvement. Digital twin technology enhances these processes by providing real-time feedback and personalized insights, allowing clinicians to adjust rehabilitation protocols effectively. Ha et al. (Ha *et al.*, 2024) introduced a hybrid upper-arm-geared exoskeleton system with an anatomical digital twin model, integrating VR torque feedback and MR remote healthcare monitoring through EMG and thermographic sensors to assess muscle activity and temperature distribution, thus ensuring safety. By simulating physical interactions, the system offers immersive training with realistic torque feedback, supporting rehabilitation through interactive virtual environments. Similarly, Ioussoufovitch and Diop (Ioussoufovitch and Diop, 2024) developed a Time-Domain Diffuse Optical Imaging system for monitoring rheumatoid arthritis in finger joints, specifically targeting the proximal interphalangeal joints. Using tissue-mimicking phantoms and spatiotemporal Fourier analysis, their system distinguishes between different RA states, enhancing monitoring precision and enabling earlier detection of treatment failures.

Together, these approaches illustrate how digital twins create a comprehensive pipeline in limb orthopedic care, from understanding anatomy and enhancing surgical precision to optimizing allograft matching and supporting effective rehabilitation, ultimately facilitating a seamless and personalized continuum of care.

### 2.4.3 Spine Vertebral Body

Spinal health is essential, with conditions like disc degeneration affecting over 70% of people under 50 and more than 90% of those over 50 (Teraguchi *et al.*, 2014). Digital twin technology plays a critical role in managing spine-related disorders by enabling accurate simulations, real-time monitoring, and personalized treatment planning. This section discusses three key applications: ReconGAN for Orthopedic Interventions, Integration with Real-Time Systems, and Digital Twins in Forensic Anthropology and Morphological Analysis.

Ahmadian et al. (Ahmadian *et al.*, 2022a) used ReconGAN to simulate vertebroplasty, crucial for vertebral stability in cancer patients, by generating realistic trabecular bone microstructures and predicting the effects of bone cement injections. In another study (Ahmadian *et al.*, 2022b), they applied ReconGAN to predict fracture risks from micro-CT images, producing high-fidelity models to simulate fracture responses. These applications demonstrate ReconGAN's ability to enhance clinical outcomes by personalizing spinal interventions and assessing risks in musculoskeletal treatment.

Significant advancements in digital twin technology have enabled real-time prediction and analysis for lumbar and cervical spine health. He et al. (He *et al.*, 2021) developed a real-time lumbar spine model integrating FEM, motion capture, IK, and machine learning. This framework uses FEM to simulate lumbar biomechanics, motion capture for movement data, and IK for accurate joint modeling. A GPR algorithm predicts biomechanical responses such as facet contact force and intradiscal pressure, visualized dynamically in Unity3D for enhanced understanding of spine mechanics.

Similarly, Moztarzadeh et al. (Moztarzadeh *et al.*, 2023) leveraged the MobileNetV2 algorithm for real-time diagnosis of cervical vertebral maturation via lateral cephalometric radiographs. Integrated within a blockchain-based metaverse platform, this lightweight algorithm ensures rapid analysis and secure data management. Together, these systems highlight the use of digital twins for accurate, real-time insights in spine diagnostics and treatment planning, enhancing patient outcomes in spine-related healthcare.

## 2.5 Breast

Studies focusing on digital twins for breast conditions primarily concentrate on breast cancer, including specific subtypes like TNBC (Lehmann *et al.*, 2011). To obtain detailed imaging data, these investigations mainly employ MRI and ultrasound-based techniques such as elastography. However, the majority of datasets remain non-public. In some instances, data access is possible through permission requests or by registration. Additionally, there is a growing trend toward integrating diverse data types—including clinical records and biological samples—to enhance the precision and practical applicability of digital twins in breast health. The computational modeling methods utilized are highly varied, spanning from traditional finite-element models and graph-based approaches to advanced machine learning algorithms and biologically inspired models.

**Table 5.** Summary of digital-twin studies for breast.

| **Reference** | **Clinical Site** | **Data Modality** | **Data Quantity** | **Data Publicity** | **Computational Models** |
|---|---|---|---|---|---|
| (Jiang *et al.*, 2014) | Breast Cancer | 3D Ultrasound | 2 Phantoms | Not Public | 3D Annotation, Spatial Sensor |
| (Sayed *et al.*, 2020) | Breast Cancer | Elastography, Ultrasound | 10 Patients | Not Public | Finite-Element Modeling |
| (Moztarzadeh *et al.*, 2023) | Breast Cancer | MRI, Blood Samples | 64 Patients | Not Public | LR, DTR, RFR, GBA |
| (Wu *et al.*, 2022) | TNBC | DCE-MRI, DW-MRI | 56 Patients | Not Public | Biologically-Based Models |
| (Moore *et al.*, 2024) | Breast Cancer | SEER Cancer Registry | 87,674 Patients | Available Upon Registration | Graph-Based Approach, MPoM |
| (Gamage *et al.*, 2023) | Breast Cancer, HF | MRI, Clinical Data | 922 Patients | Public | Biomechanics, Machine Learning |

Digital twin technology is playing a pivotal role in breast healthcare, with diverse applications in imaging, diagnosis, treatment, and clinical data integration. For instance, Jiang et al. (Jiang *et al.*, 2014) developed an innovative semi-automated 3D annotation method for breast ultrasound using a digital twin model, which significantly enhanced both the accuracy and repeatability of annotations. This work beautifully illustrates how digital twins can serve as a bridge between conventional 2D ultrasound and advanced 3D anatomical visualization, offering a more precise, dynamic representation of breast anatomy. The resulting improvements in accuracy and clinical feasibility underscore the immense value of integrating digital twins into routine breast imaging, paving the way for more standardized and efficient diagnostic procedures.

Building upon these advancements in imaging, the role of digital twins in breast cancer diagnosis is further underscored by the work of Sayed et al. (Sayed *et al.*, 2020), who applied FEM to simulate breast tissue behavior during elastography. This study demonstrated that 3D FEM could serve as a functional biomarker for distinguishing between benign and malignant tumors, providing a non-invasive, data-driven approach to early cancer detection. Such applications reinforce the practical relevance of digital twins in replicating complex biomechanical properties of breast tissues. Similarly, Moztarzadeh et al. (Moztarzadeh *et al.*, 2023) utilized multimodal data and machine learning to construct a digital twin model for breast cancer diagnosis and treatment, significantly improving diagnostic accuracy and enabling real-time disease monitoring. This work shows how digital twins can integrate diverse data sources to create dynamic, patient-specific models.

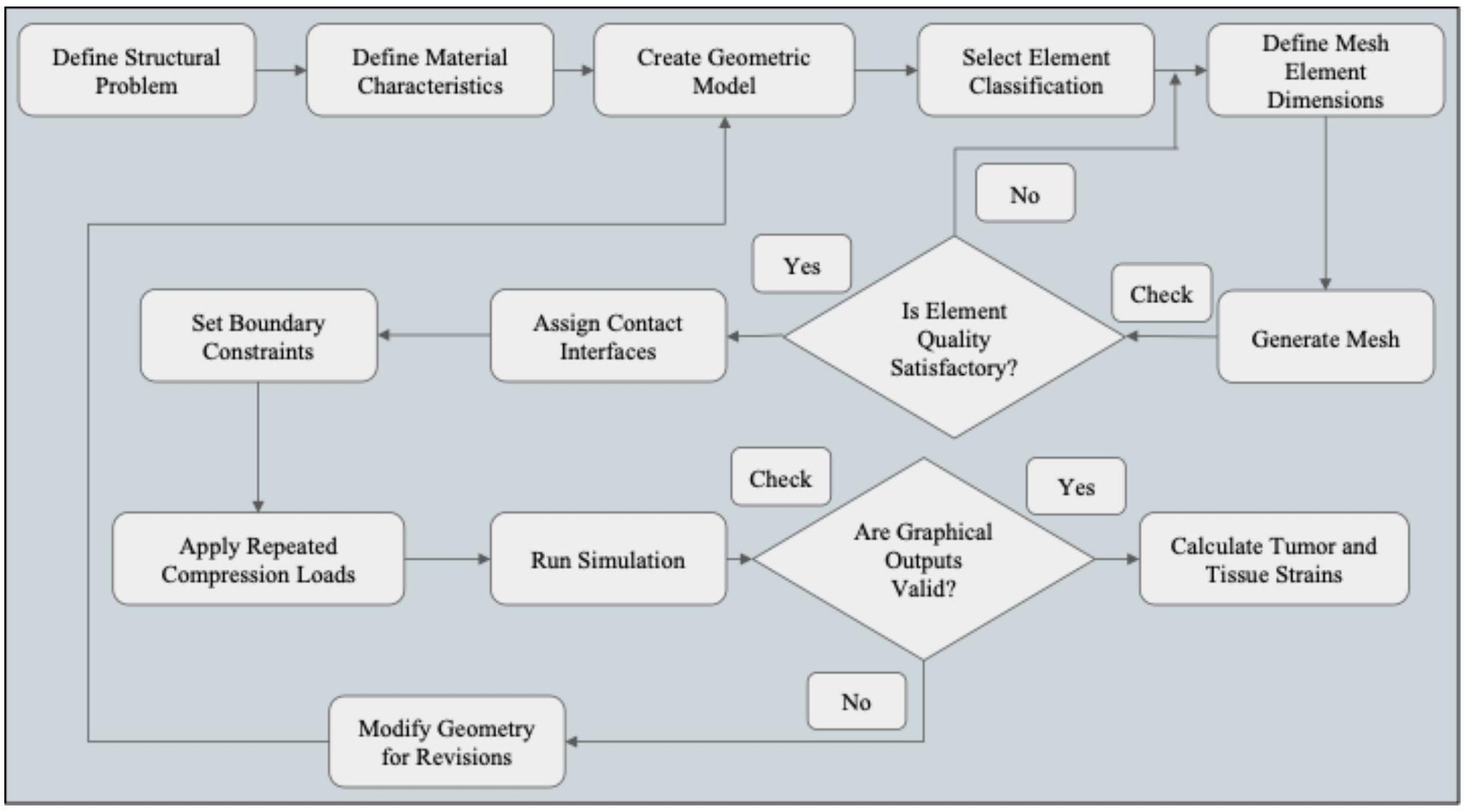

**Figure 5**. Workflow of FEM for simulating breast tissue multicompression in ultrasound elastography. Adapted from (Sayed *et al.*, 2020).

As illustrated in Figure 5, the process flow provides a detailed overview of the FEM steps used for simulating breast tissue during multicompression in ultrasound elastography. The figure outlines the comprehensive workflow, from defining the problem as a solid structure specifying material properties and building the object's geometry. The process includes selecting element type and size, meshing, and ensuring element quality. Once the quality is acceptable, boundary conditions and contact elements are applied, followed by the multi-compression application and solution phase. The graphical results are evaluated to ensure they are adequate, and if so, the tumor and normal tissue strain calculations are performed. If the results are unsatisfactory, the process revisits the geometry for adjustments. This structured approach assists in differentiating benign from malignant tumors based on their biomechanical properties.

Further advancing the use of digital twins, Moore et al. (Moore *et al.*, 2024) introduced SynTwin, a model that leverages synthetic data for clinical outcome prediction. The enhanced predictive accuracy achieved with SynTwin, particularly in breast cancer mortality prediction, demonstrates the immense value of digital twins in synthesizing clinical data to support personalized treatment strategies. Additionally, Wu et al. (Wu *et al.*, 2022) developed a biologically-based digital twin model using multiparametric MRI to predict individual responses to NAST in TNBC patients, achieving high predictive accuracy. Finally, Gamage et al. (Gamage *et al.*, 2023) introduced the 12 LABOURS Digital Twin Platform, which integrates multiple data sources to automate computational physiology workflows, thereby enhancing the efficiency of clinical trials and supporting breast cancer diagnosis and treatment.

### 2.6 Urinary System

Digital twin research focusing on the urinary tract utilizes various imaging techniques—including CT scans, MRI, and endoscopic images—to model conditions such as renal masses, kidney tumors, and prostate cancer. Besides, some researchers investigated the impact of digital twin technology in urologic oncology, specifically focusing on the treatments like RARP (Novara *et al.*, 2012). Moreover, three-dimensional virtual and printed models are widely employed, especially in surgical planning and complex anatomical reconstructions. While most datasets in this domain are not publicly accessible, a notable large-scale kidney condition dataset is available through Kaggle, demonstrating efforts to improve data availability. The computational modeling approaches used are diverse, ranging from traditional methods like 3DVM analysis to some newly-invented deep learning algorithms. This evolution reflects a growing trend toward leveraging machine learning to enhance accuracy and predictive capabilities in digital twin applications for urinary health.

**Table 6.** Summary of digital-twin studies for urinary system.

| Reference | Clinical Site | Data Modality | Data Quantity | Data Publicity | Computational Models |
|---|---|---|---|---|---|
| (Amparore *et al.*, 2024) | Renal Mass | CT, Endoscopic Images | 20 Patients | Not Public | CNN, Computer Vision |
| (Checcucci *et al.*, 2024) | Kidney Cancer | CT, MRI, 3D Virtual Models | 9 Patients | Not Public | Virtual Reality, Surgical Planning |
| (Grosso *et al.*, 2024) | Renal Mass | CT, 3D Virtual Models | 152 Patients | Not Public | Propensity-Score Matching, 3DVM Analysis |
| (Sasikaladevi and Revathi, 2024) | Kidney Stones, Cysts, Tumor, Chronic Kidney Disease | CT-Radiography | 12,446 Images | Public Dataset (Kaggle) | HGCNN, EfficientNetB0 |
| (Sarhan *et al.*, 2024) | Prostate Cancer | mp-MRI, 3D Printed Models | 1075 Patients | Not Public | 3D Virtual Models, Meta-Analysis |

### 2.6.1 Kidney

Recent advancements in digital twin technology have fundamentally transformed surgical planning, simulation, and diagnostic imaging, offering unprecedented insights and precision in kidney healthcare. Amparore and Sica (Amparore *et al.*, 2024) provide a compelling illustration of this transformation through the application of computer vision and machine learning in AR-guided robotic partial nephrectomy. Their work demonstrates the power of automatic 3D virtual image overlays in real-time surgical environments. By integrating intraoperative endoscopic images with patient-specific 3D models derived from CT scans, they showed that digital twins not only improved registration accuracy but also minimized the need for manual adjustments, thereby enhancing both surgical efficiency and precision.

In the context of preoperative surgical planning, Checcucci and Amparore (Checcucci *et al.*, 2024) employed three-dimensional virtual models and metaverse technology to enhance minimally invasive partial nephrectomy procedures. This innovative approach allowed surgeons to interact with hyperaccurate 3D models within a virtual reality environment, facilitating comprehensive preoperative discussions and optimizing clamping strategies. The integration of digital twins into the planning phase led to greater precision in surgical execution by providing a detailed visualization of patient-specific anatomy and supporting data-driven decision-making. Likewise, Grosso et al. (Grosso *et al.*, 2024) demonstrated that employing 3D virtual models during robot-assisted partial nephrectomy significantly improved functional preservation and reduced surgical complications. Furthermore, Sasikaladevi and Revathi (Sasikaladevi and Revathi, 2024) utilized CT imaging to construct a digital twin model of the renal system aimed at early disease detection. Their HGNN achieved high accuracy in classifying renal pathologies, highlighting the efficacy of digital twins as non-invasive and reliable diagnostic tools that facilitate early detection and effective management of kidney diseases.

### 2.6.2 Prostate

Recent advancements in digital twin technology have fundamentally transformed surgical planning, simulation, and diagnostic imaging, offering unprecedented insights and precision in prostate healthcare.

Sarhan et al. (Sarhan *et al.*, 2024) explored the impact of digital twin technology in urologic oncology, particularly in the context of RARP. Their systematic review and meta-analysis leveraged patient-specific mp-MRI data to create both virtual and 3D-printed prostate models. These models played a crucial role in preoperative planning and intraoperative navigation, significantly reducing positive surgical margin rates. Specifically, the group using 3D models had an 8.1% positive surgical margin rate compared to 22.2% in the conventional surgery group. This study exemplifies how digital twin technology allows for the generation of highly detailed prostate models, demonstrating its efficacy in enhancing surgical precision and minimizing complications.

## 2.7 Digestive System

Digital twin research related to the digestive system employs a broad spectrum of data modalities—including MRI, CT imaging, histopathological slides, and clinical records—to investigate conditions such as colorectal cancer, NAFLD, and type 2 diabetes. Additionally, less conventional imaging techniques like 4D-CT and holographic AR are utilized, particularly in the study of liver tumors. While most datasets are not publicly accessible or require special requests for access, some are shared under specific conditions to facilitate research. The computational models applied in these studies are diverse, ranging from traditional approaches like Monte Carlo simulations (Metropolis *et al.*, 1953) and closed-loop systems to advanced techniques such as machine learning algorithms and augmented reality integration (Azuma, 1997). This progression signifies a growing trend toward incorporating modern computational methods—such as ML and the IoT (Atzori *et al.*, 2010)—to enhance the accuracy and functionality of digital twins in digestive health.

**Table 7.** Summary of digital-twin studies for digestive system.

| Reference | Clinical Site | Data Modality | Data Quantity | Data Publicity | Computational Models |
|---|---|---|---|---|---|
| (Peshkova *et al.*, 2023) | Colorectal Cancer | Histological Slides, Clinical Data | 118 Cases | Not Public | QuPath Software, ML |
| (Schutt *et al.*, 2022) | Ascending Colon Conditions | MRI, DMP | N/A | Available Upon Request | DMP, SPH, LSM |
| (Jackson *et al.*, 2024) | IPSS | CTA, 3D Modeling | 32 Dogs | Contact Corresponding Author | Materialise Mimics, 3-matic |
| (Golse *et al.*, 2021) | Post-Hepatectomy PHT | MRI, Flowmetry, CT | 47 Patients | Available Upon Request | 0D Closed-Loop Model |
| (Shi *et al.*, 2022) | Respiratory Liver Tumor | 4D-CT, Holographic AR | 2 Dogs | Available Upon Request | CNN, AR Fusion |
| (Shrestha *et al.*, 2024) | NAFLD | MRI, Monte Carlo | 68 Samples | Available Upon Request | Single-R2*, Dual-R2*, Monte Carlo |
| (Joshi *et al.*, 2023) | Type 2 Diabetes, MAFLD | MRI-PDFF, Clinical Data | 319 Patients | Available Upon Request | ML, IoT |

### 2.7.1 Liver

Recent research underscores the transformative potential of digital twin technology in liver disease management, encompassing imaging diagnosis, surgical planning, and personalized metabolic care. Jackson et al. (Jackson *et al.*, 2024) demonstrated the clinical applicability of 3D models for IPSS in dogs, illustrating that the use of 3D digital twin models significantly enhances procedure planning by providing detailed visualization of liver anatomy. This improvement in visualization translates to enhanced accuracy and safety of interventions compared to traditional 2D imaging techniques. In a similar vein, Shrestha and Esparza (Shrestha *et al.*, 2024) employed digital twins to simulate hepatic steatosis by generating MRI signals based on fat droplet morphology and comparing different models under various conditions. Their findings emphasized the robustness of the single-R2* model for fat fraction estimation in clinical settings, highlighting the value of digital twins in providing reliable, non-invasive quantification of liver disease—a crucial component for accurate diagnosis and ongoing monitoring.

In the realm of liver surgery, Golse et al. (Golse *et al.*, 2021) employed a zero-dimensional closed-loop mathematical model to create patient-specific digital twins capable of accurately predicting surgical risks, such as postoperative portal hypertension. This capability enhances personalized surgical planning by allowing clinicians to anticipate complications and tailor interventions to the specific needs of individual patients, thereby improving surgical outcomes. Furthermore, Shi et al. (Shi *et al.*, 2022) integrated digital twin models with AR to address the challenges posed by respiratory motion during liver puncture procedures. Their approach led to significant improvements in accuracy and real-time navigation, demonstrating the practical benefits of digital twins in complex surgical environments where precision is paramount. In the context of personalized metabolic care, Joshi et al. (Joshi *et al.*, 2023) uses continuous glucose monitoring, IoT sensors, and AI algorithms to create digital twins for managing metabolic diseases, resulting in significant improvements in glycemic control and liver health outcomes compared to standard care practices.

### 2.7.2 Intestine

Digital twin technology in healthcare is significantly advancing our understanding and management of digestive system functions, particularly through innovations in colorectal cancer modeling and simulations of intestinal dynamics.

To begin with, the study by Peshkova et al. (Peshkova *et al.*, 2023) offers a compelling example of how a digital twin-based approach can revolutionize pathological analysis in colorectal cancer. Collaborating with a multidisciplinary team to evaluate and annotate samples, they developed a highly detailed digital model that integrates histological data with advanced imaging techniques. This model, and the resulting colorectal cancer case database, improved the efficiency of pathological evaluations by 50% while increasing diagnostic accuracy by 20%. These advances significantly streamlined the diagnostic process, enhancing both research capabilities and clinical decision-making.

Additionally, the work of Schutt et al. (Schutt *et al.*, 2022) represents another critical leap forward in simulating intestinal dynamics and optimizing drug delivery methods. Utilizing a DCM, the researchers created a digital twin that replicates the hydrodynamic conditions of the human colon. By employing advanced simulation techniques such as DMP and SPH, they achieved a precise simulation of flow rates and shear stresses, providing a physiologically accurate environment for testing drug dissolution. This method surpasses traditional approaches, signaling a paradigm shift in how we model and understand gastrointestinal processes.

## 2.8 Respiratory System

Research on digital twins in the respiratory field encompasses a broad spectrum of conditions, such as COVID-19, lung cancer, pneumonia, and pulmonary embolism. To investigate these diseases, primary data modalities like CT scans, MRI, chest X-rays, and various clinical data inputs are utilized. Moreover, the incorporation of IoT sensors in respiratory studies highlights the emphasis on real-time monitoring capabilities. The computational models applied are diverse, ranging from advanced deep learning architectures—such as U-Net and YOLOv8 (Terven *et al.*, 2023)—to frameworks involving mixed reality and the IoMT (Joyia *et al.*, 2017). Data availability is inconsistent; while some datasets are publicly accessible—for example, a substantial collection of chest X-ray images on Kaggle—others require special access requests. This progression indicates a growing focus on leveraging machine learning and interconnected technologies to enhance the accuracy and practical applications of digital twins in respiratory health monitoring and diagnostics.

**Table 8.** Summary of digital-twin studies for respiratory system.

| Reference | Clinical Site | Data Modality | Data Quantity | Data Publicity | Computational Models |
|---|---|---|---|---|---|
| (Tai *et al.*, 2022) | Lung Cancer, Pulmonary Embolism | Clinical Data, CT, MR | 90 Patients, 1372 Controls | Available Upon Request | rAC-GAN, IoMT, Mixed Reality |
| (Zhu *et al.*, 2023) | Lung Diseases | EIT, CT | 20 Patients | Available Upon Request | IR-Net (U-Net), DL |
| (Avanzato *et al.*, 2024) | COVID-19, Lung Opacity, Pneumonia, Tuberculosis | Chest X-Ray, IoT Sensors | 23,442 Images | Public Datasets (Kaggle) | YOLOv8, AI |
| (Li, 2022) | Pneumonia | CT | 6012 Subjects | Available Upon Request | Anchor-Free Object Detection, Multi-Branch Network |

To truly understand the transformative potential of digital twin technology in respiratory healthcare, it is essential to consider its diverse applications in lung health monitoring, cancer diagnosis, and surgical simulation. The reviewed studies serve as compelling examples of how this technology is pushing the boundaries of traditional medical practices, creating more dynamic and responsive healthcare solutions. The paper by Zhu et al. (Zhu *et al.*, 2023) introduces a novel combination of EIT, digital twin technology, and deep learning algorithms for lung monitoring. By utilizing a 3D digital twin lung model and a deep learning-based image reconstruction algorithm, the study demonstrates a significant improvement in image quality compared to traditional EIT methods, achieving an SSIM of 0.737 under noisy conditions. This improvement underscores the potential of digital twin technology in creating realistic, dynamic models of specific organs—here, the lung—that can be used for more accurate and robust diagnostic purposes.

Building on these advancements, another notable example is the work by Li (Li, 2022), which proposes an innovative framework for real-time object detection using digital twin technology. This framework leverages digital twins to create a dynamic and responsive environment for medical imaging analysis. By incorporating a five-level feature pyramid to capture fine and coarse details and employing a dual-branch detection head for accurate localization and sizing, the study demonstrates superior performance on various datasets, including COCO and pneumonia detection datasets, surpassing state-of-the-art methods like RetinaNet. This example illustrates how digital twin technology enhances object detection capabilities in medical applications, providing both higher accuracy and real-time efficiency, which are critical for timely and precise healthcare interventions. Li's contributions emphasize the versatility of digital twin technology in handling complex medical imaging tasks.

In addition to these diagnostic and imaging advancements, Avanzato et al. (Avanzato *et al.*, 2024) illustrate the application of digital twin technology for real-time lung health monitoring and disease diagnosis through the Lung-DT framework. This system integrates IoT sensors and AI algorithms to create a comprehensive digital representation of a patient's respiratory health, effectively generating a digital twin of the respiratory system that can continuously monitor health status. The YOLOv8 neural network enables accurate classification of chest X-rays into different lung disease categories with high precision and recall, achieving an average accuracy of 96.8%. This framework highlights the advantage of using digital twins for personalized and continuous lung health monitoring. This work showcases the potential of digital twins to transform personalized health monitoring into a continuous, proactive practice, ultimately contributing to better patient outcomes.

Lastly, moving towards applications in surgical practice, the study by Tai et al. (Tai *et al.*, 2022) focuses on the use of digital twin technology in remote lung cancer surgery simulation, showcasing its capacity to enhance surgical precision and immersion. By enhancing a CNN network and introducing the DL-CNN model, the study achieves high accuracy in quantifying clot burden in pulmonary embolism patients, supported by advanced haptic feedback for real-time surgical simulation. This approach leverages the digital twin to create a realistic, data-driven surgical environment, allowing for real-time visualization and tactile feedback, which vastly surpasses traditional methods. It emphasizes their potential for improving both the management of clinical data and the effectiveness of remote surgical training, making it possible to simulate complex procedures in a highly controlled and immersive manner.

## 3. Discussion

Advancing digital twin technology in the medical field with the support of medical imaging presents a highly promising direction. However, realizing its full potential requires overcoming several technical challenges. These challenges can be broadly categorized into computational performance bottlenecks, data scarcity, accuracy and validation, and data integration issues, providing a structured framework for addressing the obstacles in this field.

Data scarcity presents a critical challenge in developing digital twin models, fundamentally limiting the diversity and representativeness of the datasets available for training. This scarcity introduces biases, as models trained on narrow population samples struggle to generalize effectively, often resulting in overfitting—where a model performs well on specific training data but fails with broader, more diverse populations. For instance, similar challenges arise in musculoskeletal systems: Swaitly et al. (Swaity *et al.*, 2024) claimed that their maxillary impacted canines segmentation model encountered overfitting due to insufficient and non-diverse training data. Same in neurological modeling, limited data led Cen et al. (Cen *et al.*, 2023) and Lachinov et al. (Lachinov *et al.*, 2023) to employ simpler methodologies like spline functions and NeuralODEs instead of deep learning models to avoid overfitting, which constrained their ability to represent multiple sclerosis and Alzheimer's disease comprehensively. These adaptations, while addressing immediate data limitations, come at the expense of model precision and applicability. Thus, data scarcity compels researchers to strike a balance between feasibility and model sophistication. While this is a universal challenge in the medical imaging modeling field, positively, this challenge also drives innovation, encouraging strategies to emerge, which leads to the application of AI and ML in the construction of digital twin models. While traditional data augmentation methods involve: flip, rotate, crop, zoom and so on (Krizhevsky *et al.*, 2012), advanced methods like synthetic data using GAN (Goodfellow *et al.*, 2020) and Diffusion Model (Sohl-Dickstein *et al.*, 2015) start getting popular. This novel approach was firstly brought into medical field by Frid-Adar for liver lesion classification, and soon this method has been explored and advanced by others in medical image (Shin *et al.*, 2018; Sundaram and Hulkund, 2021; Yoon *et al.*, 2023), and physiological data (Chakshu *et al.*, 2021). Given the promising results in general medical data generation, GAN- and diffusion-based models offer valuable opportunities for future researchers to enhance digital twin construction, providing diverse and synthetic data that can help overcome limitations in real-world datasets.

Computational complexity remains a core challenge in implementing digital twin technology within medical imaging. Even with abundant data, the immense computational resources required to run these models limits their practicality in clinical settings. Digital twins aim to replicate complex physiological systems with high fidelity, which inherently demands extensive computational power—especially critical for real-time applications. For example, Lu et al. (Lu *et al.*, 2023) demonstrated that creating a digital twin of the brain required over 3,500 nodes and 10,000 GPUs, a level of demand that is impractical for most healthcare facilities. Similarly, Ahn et al. (Ahn *et al.*, 2024) faced substantial difficulties in applying FEA for biomechanical modeling due to its intense computational requirements, making real-time clinical integration exceptionally challenging. HPC and parallel processing have been adopted to address these computational demands, but they come with limitations of their own. The high costs and infrastructure requirements of HPC resources restrict their accessibility, creating a barrier to widespread digital twin implementation in diverse healthcare settings. For instance, Ahmadian et al. (Ahmadian *et al.*, 2022b) found that even with access to HPC, reconstructing vertebral microstructures was computationally prohibitive, demonstrating that simply adding computational power is not always a viable solution. These cases illustrate that computational constraints are not just technical hurdles; they are crucial issues that impact the scalability and integration of digital twins, particularly where real-time model updates are necessary to support dynamic clinical decision-making. To address these challenges, researchers are rethinking using latest ML algorithms to improve the model structures and workflow to balance fidelity with feasibility rather than relying solely on computational power. For example, in model wise, Zhou et al. (Zhou *et al.*, 2018) achieved a 32.2% reduction in inference time through model pruning, greatly decreasing UNet model depth and complexity. Similarly, Koopsen et al. (Koopsen *et al.*, 2024) reduced parameters to simplify cardiovascular models, and lowered computational demands while retaining essential accuracy. From workflow perspective, Banerjee et al. (Banerjee *et al.*, 2021) optimized their pipeline by constructing 3D heart models from 2D cine MR images, enhancing efficiency. Though not universally applicable, these strategies highlight a direction worth exploring for constructing efficient, accurate models.

Achieving precise registration and alignment is a crucial challenge in implementing digital twin DT technology in medical imaging, especially for patient-specific modeling, where anatomical accuracy is essential. This challenge is particularly evident in musculoskeletal applications, where aligning digital twin replicas with skeletal structures is complicated by the rigidity and anatomical variability of bones. Any misalignment between the digital twin and the patient's actual anatomy can lead to significant errors in diagnosis and treatment planning. During the construction phase, precise registration ensures that the digital twin accurately mirrors the patient's unique anatomical features, which allows clinicians to simulate and plan interventions with a higher degree of confidence. In dynamic contexts, such as the shoulder joint, studies by Verweij et al. (Verweij *et al.*, 2024) and Kim et al. (Kim *et al.*, 2023) emphasize the need for adaptive algorithms to manage movement variability, which is vital for real-time updates in musculoskeletal digital twins. Similarly, Vila et al. (Vila *et al.*, 2024) and Ahmadian et al. (Ahmadian *et al.*, 2022b) highlight the importance of precise alignment in procedures like bone grafting and vertebroplasty, where successful outcomes heavily rely on exact imaging data alignment. Although approaches like that of Amparore et al. (Amparore *et al.*, 2024) show improved registration by integrating intraoperative endoscopic images with CT-based 3D models, robust solutions are still required to tackle alignment challenges effectively across diverse applications.

DL advancements are emerging as promising solutions for enhancing registration accuracy in complex medical imaging tasks. Models such as Segment Anything Model 2 (Ravi *et al.*, 2024), VoxelMorph (Balakrishnan *et al.*, 2019), nnU-Net (Isensee *et al.*, 2021), and DeepReg (Ledesma-Dominguez *et al.*, 2024) have shown considerable success in aligning multimodal datasets like MRI-to-CT and in applications like tumor tracking and soft tissue registration. These DL models leverage advanced learning techniques to manage the inherent variability in medical imaging, making them more adaptable than traditional registration methods. Although not yet widely applied to skeletal digital twins, the demonstrated adaptability and precision of these models suggest they could help bridge current gaps, particularly in managing the rigid and intricate nature of bone structures in digital twin applications.

In summary, while implementing digital twin technology in medical imaging presents numerous challenges—from data scarcity and computational complexity to registration—each obstacle also serves as an impetus for innovation. As we navigate these complexities, the advancements made bring us closer to the ultimate goal of improving personalized medicine and patient outcomes, demonstrating the vast potential of digital twin technology in healthcare.

## 4. Conclusion

The exploration of medical imaging applications within digital twin technology reveals its transformative potential across numerous healthcare fields. Through a comprehensive analysis of advancements, challenges, and emerging research, this review illustrates that digital twins represent more than mere technological innovation; they are pivotal tools poised to revolutionize personalized medicine. By integrating research across cardiovascular, neurological, musculoskeletal, and other systems, we demonstrate the profound impact of digital twins in fostering a future where predictive, precision-based healthcare becomes the standard. These virtual replicas, formed by combining high-resolution imaging with advanced computational models, enable patient-specific treatments and diagnostics. Despite challenges such as data scarcity and computational complexity, these obstacles have catalyzed innovative solutions, leading to practical clinical applications like optimized cardiovascular care, neurological disease modeling, and precision-guided musculoskeletal surgeries. Looking forward, the ongoing development of digital twin technologies holds the promise of a proactive, personalized healthcare future, driving enhanced patient outcomes and advancing the field of medicine. This review, therefore, serves as a guide for researchers and clinicians, encouraging them to further harness the transformative potential of digital twin technology in healthcare.

**Acknowledgements**

This research is supported in part by the National Institutes of Health under Award Number R01CA272991, R56EB033332, and R01DE033512.